\documentclass[journal]{IEEEtran}
\usepackage{graphicx}
\usepackage{amssymb,amsmath}
\usepackage{bm}

\usepackage[english]{babel}
\usepackage[T1]{fontenc}
\usepackage[utf8]{inputenc}

\usepackage{hyperref}
\hypersetup{hidelinks,
  pdftitle={{Deep Learning for Audio Signal Processing}},
  pdfauthor={{Hendrik Purwins, Bo Li, Tuomas Virtanen, Jan Schlüter, Shuo-yiin Chang, Tara Sainath}},
}

\usepackage{cite}

\ifCLASSINFOpdf
\else
\fi
\begin{document}

\title{Deep Learning for Audio Signal Processing}
%
%
%

\author{Hendrik Purwins$^*$, Bo Li$^*$, Tuomas Virtanen$^*$, Jan Schl{\"u}ter$^*$, Shuo-yiin Chang, Tara Sainath 
        
\thanks{$^*$ H.\ Purwins, B.\ Li, T.\ Virtanen, and J.\ Schl{\"u}ter contributed equally to this paper.}        
\thanks{H.\ Purwins is with Department of Architecture, Design \& Media Technology, Aalborg University Copenhagen, e-mail: hpurwins@gmail.com.}
\thanks{B.\ Li, S.-y.\ Chang and T.\ Sainath are with Google, e-mail: \{boboli,shuoyiin,tsainath\}@google.com}
\thanks{T.\ Virtanen is with  Tampere University, e-mail: tuomas.virtanen@tuni.fi}
\thanks{J.\ Schl{\"u}ter is with CNRS LIS, Université de Toulon and Austrian Research Institute for Artificial Intelligence, e-mail: jan.schlueter@ofai.at.}
\thanks{Great thanks to Duncan Blythe for proof-reading.}
\thanks{Manuscript received October 11, 2018
}
\thanks{This is a PREPRINT}
}

%
%

\markboth{Journal of Selected Topics of Signal Processing,~Vol.~13, No.~2, May~2019, pp.~206--219}%
{Shell \MakeLowercase{\textit{et al.}}: Deep Learning for Audio Signal Processing}
%



\setcounter{page}{0}
\thispagestyle{empty}
\begin{figure*}[p!]
\centering
\parbox[t][.8\textheight][t]{1.5\columnwidth}{\large
\copyright 2019 IEEE.  Personal use of this material is permitted. Permission from IEEE must be obtained for all other uses, in any current or future media, including reprinting/republishing this material for advertising or promotional purposes, creating new collective works, for resale or redistribution to servers or lists, or reuse of any copyrighted component of this work in other works.

\vspace{1em}
Appeared as:
Hendrik Purwins, Bo Li, Tuomas Virtanen, Jan Schlüter, Shuo-yiin Chang, Tara Sainath.
Deep Learning for Audio Signal Processing.
In \emph{Journal of Selected Topics of Signal Processing},~Vol.~13, No.~2, May~2019, pages~206--219.

\vspace{1em}
\url{http://doi.org/10.1109/JSTSP.2019.2908700}
}
\end{figure*} 
\maketitle

\begin{abstract}
Given the recent surge in developments of deep learning, this article provides a review of the state-of-the-art deep learning techniques for audio signal processing. Speech, music, and environmental sound processing are considered side-by-side, in order to point out similarities and differences between the domains, highlighting general methods, problems, key references, and potential for cross-fertilization between areas. 
The dominant feature representations (in particular, log-mel spectra and raw waveform) and deep learning models are reviewed, including convolutional neural networks, variants of the long short-term memory architecture, as well as  more audio-specific neural network models. Subsequently, prominent  
 deep learning application areas are covered, i.e. audio recognition (automatic speech recognition, music information retrieval, environmental sound detection,  localization and tracking) and synthesis and transformation (source separation, audio enhancement, generative models for speech, sound, and music synthesis). Finally, key issues and future questions regarding deep learning applied to audio signal processing are identified.  \end{abstract}

\begin{IEEEkeywords}.
deep learning, connectionist temporal memory, automatic speech recognition, music information retrieval, source separation, audio enhancement, environmental sounds  
\end{IEEEkeywords}

\IEEEpeerreviewmaketitle

\section{Introduction} 
\IEEEPARstart{A}{rtificial} neural networks have gained widespread attention in three waves so far, triggered by  1) the  perceptron algorithm \cite{rosenblatt_perceptron1958} in 1957, 2) the backpropagation algorithm \cite{rumelhart_learning_1986} in 1986, and finally 3) the success of deep learning in speech recognition \cite{hinton2012deep} and image classification \cite{krizhevsky_imagenet_2012} in 2012, leading to a renaissance of deep learning, involving e.g.\ deep feedforward neural networks \cite{hinton2012deep,mohamed2009deep}, convolutional neural networks (CNNs, \cite{lecun_backpropagation_1989}) and long short-term memory (LSTM, \cite{hochreiter_long_1997}).
 In this "deep" paradigm, architectures with a large number of parameters are trained to learn from a massive amount of data leveraging recent advances in machine parallelism (e.g.\ cloud computing, GPUs or TPUs \cite{jouppi2017datacenter}).   
The recent surge in interest in deep learning has enabled practical applications in many areas of signal processing, often outperforming traditional signal processing on a large scale. 
In this most recent wave, deep learning first gained traction in image processing \cite{krizhevsky_imagenet_2012}, but was then widely adopted in speech processing, music and environmental sound processing, as well as numerous additional fields such as genomics, quantum chemistry, drug discovery, natural language processing and recommendation systems.  
As a result, previously used methods in audio signal processing, such as Gaussian mixture models, hidden Markov models and non-negative matrix factorization, have often  been  outperformed by deep learning models, in applications where sufficient data is available.

While many deep learning methods have been adopted from image processing, there are important differences between the domains that warrant a specific look at audio.
Raw audio samples form a one-dimensional time series signal, which is fundamentally different from two-dimensional images. Audio signals are commonly transformed into two-dimensional time-frequency representations for processing, but the two axes, time and frequency, are not homogeneous as horizontal and vertical axes in an image. Images are instantaneous snapshots of a target and often analyzed as a whole or in patches with little order constraints; however audio signals have to be studied sequentially in chronological order.
These properties gave rise to audio-specific solutions.

\section{Methods}
To set the stage,
we give a conceptual overview of audio analysis and synthesis problems (\ref{subsec:problems}), the input representations commonly used to address them (\ref{subsec:features}), and the models shared between different application fields (\ref{subsec:models}).
We will then briefly look at data (\ref{subsec:data}) and evaluation methods (\ref{subsec:evaluation}).

\subsection{Problem Categorization} 
\label{subsec:problems}

\begin{figure}
\includegraphics{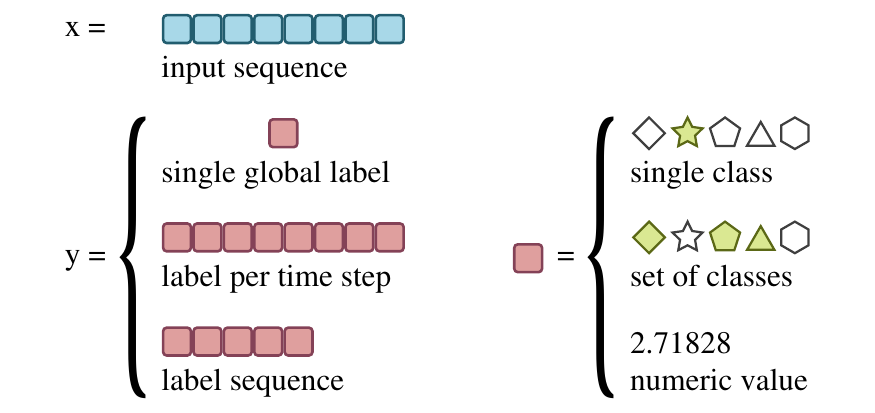}
\caption{Audio signal analysis tasks can be categorized along two properties:
The number of labels to be predicted (\emph{left}),
and the type of each label (\emph{right}).}
\label{fig:problems}
\end{figure}

The tasks considered in this survey can be divided into different categories depending on the kind of target to be predicted from the input, which is always a time series of audio samples.%
\footnote{While the audio signal will often be processed into a sequence of features, we consider this part of the solution, not of the task.}
This division encompasses two independent axes (cf.\ Fig.~\ref{fig:problems}):
For one, the target can either be a single global label, a local label per time step, or a free-length sequence of labels (i.e., of a length that is not a function of the input length).
Secondly, each label can be a single class, a set of classes, or a numeric value.
In the following, we will name and give examples for the different combinations considered.

Predicting a single global class label is termed \emph{sequence classification}. Such a class label can be a predicted language, speaker, musical key or  acoustic scene, taken from a predefined set of possible classes. 
In \emph{multi-label sequence classification}, the target is  a subset of the set of possible classes. For example, the target can  comprise several acoustical events, such as in the weakly-labelled AudioSet dataset \cite{audioset},  or a set of musical pitches.  Multi-label classification can be particularly efficient when classes depend on each other.
In \emph{sequence regression}, the target is a value from a continuous range. Estimating musical tempo or predicting the next audio sample can be formulated as such. Note that regression problems can always be discretized and turned into classification problems: e.g., when the audio sample is quantized into 8 bits, predicting the sample poses a classification problem with 256 classes.

When predicting a label per time step, each time step can encompass a constant number of audio samples, so the target sequence length is a fraction of the input sequence length. Again, we can distinguish different cases.
Classification per time step is referred to as \emph{sequence labeling} here. Examples are chord annotation and vocal activity detection.
\emph{Event detection} aims to predict time points of event occurrences, such as speaker changes or note onsets, which can be formulated as a binary sequence labeling task: at each step, distinguish presence and absence of the event.
Regression per time step generates continuous predictions, which may be the distance to a moving sound source or the pitch of a voice, or source separation.

In \emph{sequence transduction}, the length of the target sequence is not a function of the input length.
There are no established terms to distinguish classification, multi-label classification and regression.
Examples comprise speech-to-text, music transcription, or language translation. 

Finally, we also consider some tasks that do not start from an audio signal:
Audio synthesis can be cast as a sequence transduction or regression task that predicts audio samples from a sequence of conditional variables.
Audio similarity estimation is a regression problem where a continuous value is assigned to a pair of audio signals of possibly different length.

\subsection{Audio Features}\label{subsec:features}

Building an appropriate feature representation and designing an appropriate classifier for these features have often been treated as separate problems in audio processing. One drawback of this approach is that the designed features might not be optimal for the classification objective at hand. Deep neural networks (DNNs) can be thought of as performing feature extraction jointly with objective optimization such as classification. For example, for speech recognition, Mohamed et al. \cite{ar:tSNE} showed that the activations at lower layers of DNNs can be thought of as speaker-adapted features, while the activations of the upper layers of DNNs can be thought of as performing class-based discrimination.

For decades, mel frequency cepstral coefficients (MFCCs) \cite{furui1986speaker} have been used as the dominant acoustic feature representation for audio analysis tasks.
These are magnitude spectra projected to a reduced set of frequency bands, converted to logarithmic magnitudes, and approximately whitened and compressed with a discrete cosine transform (DCT).
With deep learning models, the latter has been shown to be unnecessary or unwanted, since it removes information and destroys spatial relations.
Omitting it yields the \emph{log-mel spectrum}, a popular feature across audio domains.

The mel filter bank for projecting frequencies is inspired by the human auditory system and physiological findings on speech perception \cite{davis:mel}.
For some tasks, it is preferable to use a representation which captures transpositions as translations. Transposing a tone consists of scaling the base frequency and overtones by a common factor, which becomes a shift in a logarithmic frequency scale.
The \emph{constant-Q spectrum} achieves such a frequency scale with a suitable filter bank \cite{purwins_new_2000}.

A (log-mel, or constant-Q) \emph{spectrogram} is a temporal sequence of spectra.
As in natural images, the neighboring spectrogram bins of natural sounds in time and frequency are correlated.
However, due to the physics of sound production, there are additional correlations for frequencies that are multiples of the same base frequency (harmonics).
To allow a spatially local model (e.g., a CNN) to take these into account, a third dimension can be added that directly yields the magnitudes of the harmonic series \cite{Lostanlen2016_instrumentcnn,Bittner2018_harmoniccqt}. 
Furthermore, in contrast to images, value distributions differ significantly between frequency bands.
To counter this, spectrograms can be standardized separately per band.

The window size for computing spectra trades temporal resolution (short windows) against frequential resolution (long windows).
Both for log-mel and constant-Q spectra, it is possible to use shorter windows for higher frequencies, but this results in inhomogeneously blurred spectrograms unsuitable for spatially local models. Alternatives include computing spectra with different window lengths, projected down to the same frequency bands, and treated as separate channels \cite{Schlueter2014_icassp}. In \cite{chen2014feature} the authors also investigated combinations of different spectral features.

To avoid relying on a designed filter bank, various methods have been proposed to further simplify the feature extraction process and defer it to data-driven statistical model learning. Instead of mel-spaced triangular filters, data-driven filters have been learned and used.
\cite{sainath_learning_2013} and \cite{Cakir2016_stftfilterbank} use a full-resolution magnitude spectrum,
\cite{Jaitly11,Collobert14,Zoltan14,Yedid15} directly use a \emph{raw waveform} representation of the audio signals as inputs and learn data-driven filters jointly with the rest of the network for the target tasks. In this way, the learned filters are directly optimized for the target objective in mind.
In \cite{Sainath15b}, the lower layers of the model are designed to mimic the log-mel spectrum computation but with all the filter parameters learned from the data.
In \cite{oord2016wavenet}, the notion of a filter bank is discarded, learning a causal regression model of the time-domain waveform samples without any human prior knowledge. 

\subsection{Models}
\label{subsec:models}
The audio signal, represented as a sequence of either frames of raw audio or human engineered feature vectors (e.g. log-mel/constant-Q/complex spectra), matrices (e.g. spectrograms), or tensors (e.g. stacked spectrograms), can be analyzed by various deep learning models.
Similar to other domains like image processing, for audio, multiple feedforward, convolutional, and recurrent (e.g. LSTM) layers are usually stacked to increase the modeling capability. A \emph{deep neural network} is a neural network with many stacked layers \cite{goodfellow_deep_2016}.  

\paragraph{Convolutional Neural Networks (CNNs)}
CNNs are based on convolving their input with learnable kernels.
In the case of spectral input features, a 1-d temporal convolution or a 2-d time-frequency convolution is commonly adopted, whereas a time-domain 1-d convolution is applied for raw waveform inputs. A convolutional layer typically computes multiple feature maps (\emph{channels}), each from its corresponding kernel. Pooling layers added on top of these convolutional layers can be used to downsample the learned feature maps.  
A CNN often consists of a series of convolutional layers interleaved with pooling layers, followed by one or more dense layers. 
For sequence labeling, the dense layers can be omitted to obtain a fully-convolutional network (FCN).

The receptive field (the number of samples or spectra involved in computing a prediction) of a CNN is fixed by its architecture.
It can be increased by using larger kernels or stacking more layers. Especially for raw waveform inputs with a high sample rate, reaching a sufficient receptive field size may result in a large number of parameters of the CNN and high computational complexity. 
Alternatively, a dilated convolution (also called \textit{atrous}, or convolution with holes) \cite{oord2016wavenet,holschneider1989wavelets,chen2016deeplab,yu2015multi} can be used, which applies the convolutional filter over an area larger than its filter length by inserting zeros between filter coefficients. A stack of dilated convolutions enables networks to obtain very large receptive fields with just a few layers, while preserving the input resolution 
as well as computational efficiency.

Operational and validated theories on how to determine the optimal CNN architecture  (size of kernels, pooling and feature maps, number of channels and consecutive layers) for a given task are not available at the time of writing (see also \cite{watanabe_algebraic_2009}). Currently therefore, the architecture of a CNN is largely chosen experimentally based on a validation error, which has led to some rule-of-thumb guidelines, such as fewer parameters for less data \cite{ronneberger2015u}, increasing channel numbers with decreasing sizes of feature maps in subsequent convolutional layers, considering the necessary size of temporal context, and task-related design (e.g.\ analysis or synthesis/transformation). 

\paragraph{Recurrent Neural Networks (RNNs)}
The effective context size that can be modeled by CNNs is limited, even when using dilated convolutions. 
RNNs follow a different approach for modeling sequences \cite{elman1990finding}: They compute the output for a time step from both the input at that step and their hidden state at the previous step. This inherently models the temporal dependency in the inputs, and allows the receptive field to extend indefinitely into the past.
For offline applications, \emph{bidirectional RNNs} employ a second recurrence in reverse order, extending the receptive field into the future.
In contrast to conventional HMMs, with linear growth of the number of recurrent hidden units in RNNs with all-to-all kernels, the number of representable states grows exponentially, whereas  training or inference time grows only quadratically at most \cite{lipton_critical_2015}. RNNs can suffer from vanishing/exploding gradients during training.
Many variations have been developed to address this. Long short term memory (LSTM) \cite{hochreiter_long_1997} utilizes a gating mechanism and memory cells to mitigate the information flow and alleviate gradient problems. Stacking of recurrent layers \cite{mehri_samplernn2016} and sparse recurrent networks \cite{kalchbrenner_efficient_2018}  have been found useful in audio synthesis.

Besides the use for modeling temporal sequences, LSTMs 
have been extended to model audio signals across both time and frequency domains. Frequency LSTMs (F-LSTM) \cite{Jinyu15} and Time-Frequency LSTMs (TF-LSTM) \cite{Graves07, Jinyu16, tsainath2016_tflstm}
have been introduced as alternatives to CNNs to model correlations in frequency.
Distinctly from CNNs, F-LSTMs capture translational invariance through local filters and recurrent connections. They do not require pooling operations and are more adaptable to a range of types of input features.
TF-LSTMs are unrolled across both time and frequency, and may be used to model both spectral and temporal variations through local filters and recurrent connections. TF-LSTMs outperform CNNs on certain tasks \cite{tsainath2016_tflstm}, but are less parallelizable and therefore slower.

\begin{figure}[ht]
\begin{center}
\includegraphics{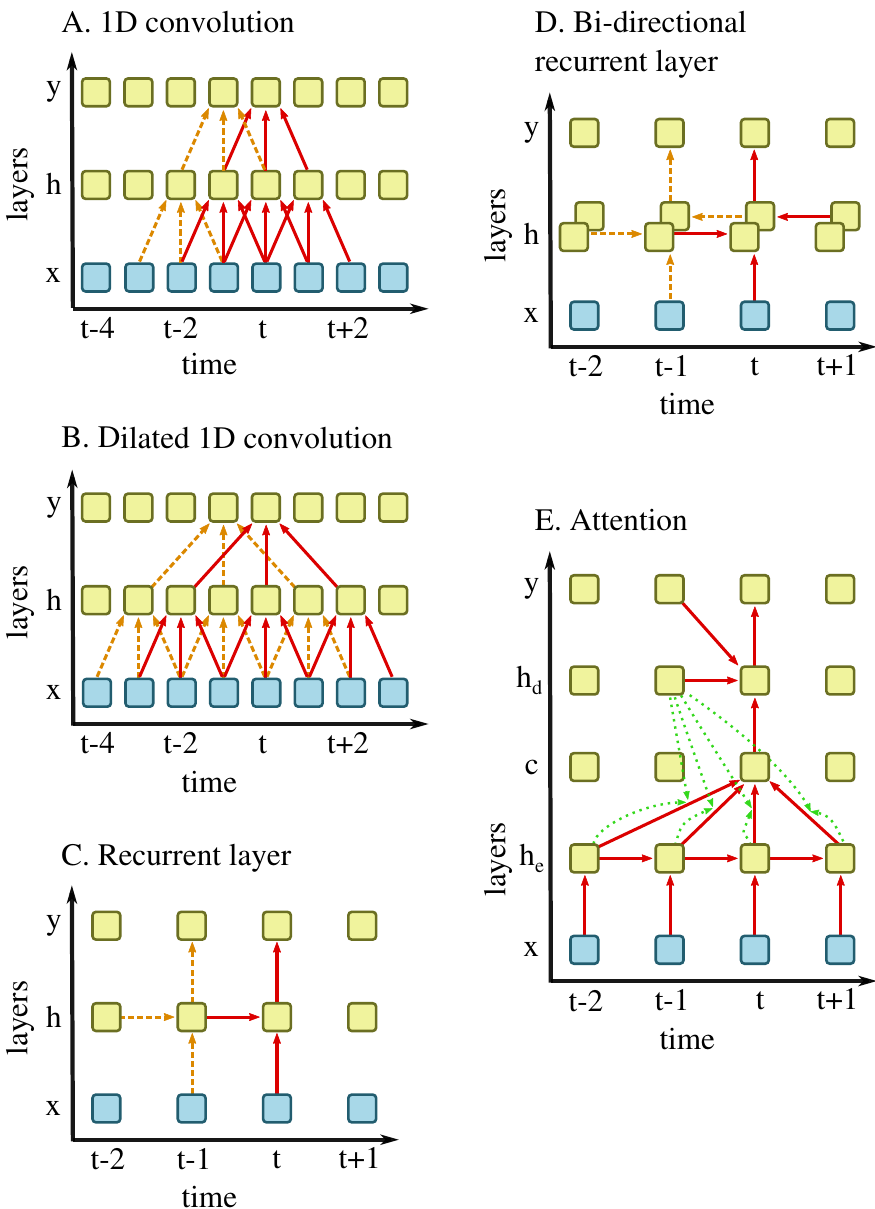}
\caption{Different ways of processing temporal context. Building blocks are shown that process an input time series $x$ via an intermediate representation $h$ into an output time series $y$. Orange dashed lines indicate processing performed for calculating output $y_{t-1}$, red solid lines mark processing yielding $y_t$. 
{\bf A.} In a convolutional layer, the representation in a layer ($h$ and $y$) is generated by convolving the activations of the previous layer with a 1-D filter, in this case consisting of 3 weights.
{\bf B.} In a dilated convolution, only every $k$th activation is taken into account, for a chosen dilation factor $k$. In this case, the second convolution is dilated by 2, so only $h_{t-2},h_{t},h_{t+2}$ are used for calculating $y_t$. However, the skipped values participate in the computation of $y_{t-1}$. Dilated convolutions can be stacked with successively increasing factors (1, 2, 4, \dots) to increase the range of the analyzed temporal context.
{\bf C.} 
In RNNs (such as GRU, LSTM), the activations in $h_t$ are calculated from the  current input $x_t$ and from previous activations  $h_{t-1}$. 
{\bf D.} In a bi-directional recurrent layer, activations in $h$ are calculated in both directions, from beginning to end and vice versa.
{\bf E.} Attention \cite{bahdanau2014neural} can be used for sequence transduction. Encoder and decoder of the network include a recurrent layer respectively as an embedding $h_e$ of the input $x$ and an embedding $h_d$ of output $y.$ The context $c_t$ is a weighted sum of the encoder embedding $h_{e,t-2},h_{e,t-1},h_{e,t},h_{e,t+1},$ where the weights are calculated  between the decoder embedding $h_{d,t-1}$  and all encoder embeddings respectively, indicated by green dotted lines. The output $y_t$ is calculated from the previous output $y_{t-1}$, the previous decoder embedding $h_{d,t-1}$ and the context $c_t$, indicating correlations between input and output positions.
}
\label{fig:context}
\end{center}
\vspace{-0.3in}
\end{figure}

Alternatively, RNNs can process the output of a CNN, forming a \emph{Convolutional Recurrent Neural Network (CRNN)}.
In this case, convolutional layers extract local information, and recurrent layers combine it over a longer temporal context. Various ways to process temporal context are visualized in Fig.~ \ref{fig:context}.

\paragraph{Sequence-to-Sequence Models}
A sequence-to-sequence model transduces an input sequence into an output sequence directly.
Many audio processing tasks are essentially sequence-to-sequence transduction tasks. However, due to the large 
complexity involved in audio processing tasks, conventional systems usually divide the task into series of sub-tasks and solve each task independently. Taking speech recognition as an example, the ultimate task entails converting the input temporal audio signals into the output sequence of words. But traditional ASR systems comprise separate acoustic, pronunciation, and language modeling components that are normally trained independently~\cite{SakSeniorBeaufays14,SainathVinyalsSeniorEtAl15}. 

With the larger modeling capacity of deep learning models, there has been growing interest in building end-to-end trained systems that directly map the input audio signal to the target sequences~\cite{Graves12,Chan15,ChorowskiBahdanauSerdyukEtAl15,SoltauLiaoSak16,ZhangChanJaitly16,LuZhangRenals16}. 
These systems are trained to optimize criteria that are related to the final evaluation metric (such as word error rate for ASR systems). Such sequence-to-sequence models are fully neural, and do not use finite state transducers, a lexicon, or text normalization modules. The acoustic, pronunciation, and language modeling components are trained jointly in a single system. This greatly simplifies training compared to conventional systems: it does not require bootstrapping from decision trees or time alignments generated from a separate system. Furthermore, since the models are trained to directly predict target sequences, the process of decoding is also simplified.

One such model is the \emph{connectionist temporal classification} (CTC). This model introduces a blank symbol to match the output sequence length with the input sequence and integrates over all possible ways of inserting blanks to jointly optimize the output sequence instead of each individual output label \cite{GravesFernandezGomezEtAl06,GravesJaitly14,ZweigYuDroppoEtAl17,Roman2018_ctctranscription}. 
The basic CTC model was extended by Graves~\cite{Graves12} to include a separate recurrent language model component, referred to as the recurrent neural network transducer (RNN-T). Attention-based models which learn alignments between the input and output sequences jointly with the target optimization have become increasingly popular~\cite{bahdanau2014neural,Chan15,vaswani2017attention}.
Among various sequence-to-sequence models, \emph{listen, attend and spell} (LAS) 
offered improvements over others~\cite{RohitSeq17} (see also Fig.~\ref{fig:context}).

\paragraph{Generative Adversarial Networks (GANs)}
GANs are unsupervised generative models that learn to produce realistic samples of a given dataset from low-dimensional, random latent vectors \cite{goodfellow2014generative}. GANs consist of two networks, a generator and a discriminator. The generator maps latent vectors drawn from some known prior to samples and the discriminator is tasked with determining if a given sample is real or fake. The two models are pitted against each other in an adversarial framework. Despite the success of GANs~\cite{goodfellow2014generative} for image synthesis, their use in the audio domain has been limited. GANs have been used for source separation \cite{subakan_generative_2018}, music instrument transformation \cite{huang2018timbretron} and speech enhancement to transform noisy speech input to denoised versions \cite{pascual2017segan,donahue2018exploring,michelsanti2017conditional,mimura2017cross}, which will be discussed in Section~\ref{sec:audio_enhancement}.

\paragraph{Loss Functions}
\label{subsec:discussion:loss}
A crucial and creative part of the design of a deep learning system is the choice of the loss function. The loss function needs to be differentiable with respect to trainable parameters of the system when gradient descent is used for training. The mean squared error (MSE) between log-mel spectra 
can be used to quantify the difference between two frames of audio in terms of their spectral envelopes. To account for the temporal structure, log-mel spectrograms can be compared. However, comparing two audio signals by taking the MSE between the samples in the time domain is not a robust measure. For example, the loss for two sinusoidal signals with the same frequency would entirely depend on the difference between their phases. To account for the fact that slightly  non-linearly warped signals sound similar, differentiable dynamic time warping distance \cite{cuturi_soft-dtw:_2017} or 
earth mover's distance such as in Wasserstein GANs  \cite{arjovsky_wasserstein_2017} might be more suitable.
The loss function can be also tailored towards particular applications. E.g. in source separation an objective  differentiable loss function can be designed based on  psychoacoustic speech intelligibility experiments. Different loss functions can be combined. For controlled audio synthesis \cite{donahue_piano_2018}, one loss function was customized to encourage the latent variables of a variational autoencoder (VAE) to remain inside a  defined range and another to have changes in the control space be reflected in the generated audio.

\paragraph{Phase modeling} 
In the calculation of the  log-mel spectrum, the magnitude spectrum is used but the phase spectrum is lost.
While this may be desired for analysis, synthesis requires plausible phases.
The phase can be estimated from the magnitude spectrum using the Griffin-Lim Algorithm \cite{daniel_w._griffin_signal_1984}. But the accuracy of the estimated phase is insufficient to yield high quality audio, desired in applications such as in source separation, audio enhancement, or generation.
A neural network (e.g. WaveNet \cite{oord2016wavenet}) can be trained to generate a time-domain signal from log-mel spectra \cite{shen2017natural}.
Alternatively, deep learning architectures may be trained to ingest the complex spectrum directly by including both magnitude and phase spectrum as input features  \cite{wilmanski_complex_2016} or via complex targets \cite{williamson2016complex}; alternatively all operations (convolution, pooling, activation functions) in a DNN may be extended to the complex domain \cite{trabelsi_deep_2017}.

When using raw waveform as input representation, for an analysis task, one of the difficulties is that perceptually and semantically identical sounds may appear at distinct phase shifts, so using a representation that is invariant to small phase shifts is critical. To achieve phase invariance researchers have usually used convolutional layers which pool in time \cite{Jaitly11,Collobert14,Yedid15} or DNN layers with large, potentially overcomplete, hidden units \cite{Zoltan14},
which are able to capture the same filter shape at a variety of phases. 
Raw audio as input representation is often used in synthesis tasks, e.g. when autoregressive models are used \cite{oord2016wavenet}.

\subsection{Data}
\label{subsec:data}
Deep learning is known to be most profitable when applied to large training datasets. For the break-through of deep learning in computer vision, the availability of ImageNet \cite{imagenet}, a database of 14 million (2019) hand-labeled images,  was a major factor. 
However, there is no such a well labeled dataset that can be shared across domains including speech, music, and environmental sounds.
For speech recognition, there are large datasets \cite{ldc}, for English in particular. 
For music sequence classification or music similarity, there is the Million Song Dataset \cite{millionsong}, whereas MusicNet \cite{musicnet} addresses note-by-note sequence labeling. Datasets for higher-level musical sequence labeling, such as chord, beat, or structural analysis are often much smaller \cite{beatles}. For environmental sound sequence classification, the AudioSet \cite{audioset} of more than 2 million audio snippets is available.

Especially in image processing, tasks with limited labeled data are solved with \emph{transfer learning}: using large amounts of similar data labeled for another task and adapting the knowledge learned from it to the target domain. For example, deep neural networks trained on the ImageNet dataset can be adapted to other classification problems using small amounts of task-specific data by \emph{retraining} the last layers or \emph{finetuning} the weights with a small learning rate. In speech recognition, a model can be pretrained on languages with more transcribed data and then adapted to a low-resource language \cite{li2018bytes} or domains \cite{narayanan2018toward}.

\emph{Data generation} and data augmentation are other ways of addressing the limited training data problem.
For some tasks, data resembling real data can be generated,  with known synthesis parameters and labels. A controlled gradual increase in  complexity of the generated data eases understanding, debugging, and improving of machine learning methods. However, the performance of an algorithm on real data may be poor if trained on generated data only. 
\emph{Data augmentation} 
generates additional training data by manipulating existing examples to cover a wider range of possible inputs. 
For ASR, \cite{Jaitly2013_vltp} and \cite{Kanda2013_elastic} independently proposed to transform speech excerpts by pitch shifting (termed \emph{vocal tract perturbation}) and time stretching. For far-field ASR, single-channel speech data can be passed through room simulators to generate multi-channel noisy and reverberant speech \cite{chanwoosim2017}.
Pitch shifting has also been shown useful for chord recognition \cite{Humphrey2012_cnn}, and combined with time stretching and spectral filtering for singing voice detection \cite{Schlueter-ismir-2015} and instrument recognition \cite{mcfee-ismir-2015}.
For environmental sounds, linearly combining training examples along with their labels improves generalization \cite{Tokozume2017_inbetween}.
For source separation, models can be trained successfully using datasets that are synthesized by mixing separated tracks.

\subsection{Evaluation} 
\label{subsec:evaluation}

Evaluation criteria vary across tasks.
For speech recognition systems, the performance is usually evaluated with word error rates (WER). WER counts the fraction of word errors after aligning the reference and hypothesis word strings and consists of insertion, deletion and substitution rates which are the number of insertions, deletions and substitutions divided by the number of reference words.
Both in music and in acoustic scene classification, accuracy is a commonly used metric.
To evaluate binary classification without a fixed classification threshold, the area under the receiver operating characteristic curve (AUROC) is an alternative to accuracy as a performance metric. The design of a performance metric may take into account semantic relationships between the classes. E.g., the loss for a chord detection task can be designed to be smaller if the detected and the actual chord are  harmonically closely related.  In event detection, performance is typically measured using equal error rate or F-score, where the true positives, false positives and false negatives are calculated either in fixed-length segments or per event \cite{Lacoste2006_mlp,mesaros-aa-2016}.
Objective source separation quality is typically measured with metrics such as signal-to-distortion ratio, signal-to-interference ratio, and signal-to-artifacts ratio \cite{vincent-taslp-2006}. 
The mean opinion score (MOS) is a subjective test for evaluating  quality of synthesized audio, in particular speech. A Turing test can also provide an evaluation measure for audio generation.

\section{Applications}
To lay the foundation for cross-domain comparisons, we will now look at concrete applications of the methods discussed, first for analyzing speech (Sec.~\ref{subsec:analysis:speech}), music (Sec.~\ref{subsec:analysis:music}) and environmental sound (Sec.~\ref{subsec:analysis:environmental}), and then for synthesis and transformation of audio: source separation (Sec.~\ref{subsec:synthesis:separation}), speech enhancement (Sec.~\ref{subsec:synthesis:enhancement}), and audio generation (Sec.~\ref{subsec:synthesis:generative}).
\subsection{Analysis}
\label{subsec:analysis}
\subsubsection{Speech} 
\label{subsec:analysis:speech}
Using voice to access information and to interact with the environment is a deeply entrenched and instinctive 
form of communication 
for humans. Speech recognition -- converting speech audio into sequences of words -- is a prerequisite to any speech-based interaction. Efforts in building automatic speech recognition systems date back more than half a century \cite{juang2005automatic}. However the vast adoption of such systems in real-world applications has only occurred in the recent years. 

For decades, the triphone-state Gaussian mixture model (GMM) / hidden Markov model (HMM) was the dominant choice for modeling speech. These models have many advantages, including their mathematical elegance, which leads to many principled solutions to practical problems such as speaker or task adaptation. 
Around 1990, discriminative training was found to yield better performance than models trained using maximum likelihood. Neural network based hybrid models were proposed to replace GMMs \cite{waibel1990phoneme,robinson1996use,bourlard2012connectionist}. However, recently in 2012, DNNs with millions of parameters trained on thousands of hours of data were shown to reduce the word error rate (WER) dramatically on various speech recognition tasks \cite{hinton2012deep}. In addition to the great success of deep feedforward and convolutional networks \cite{sainath13}, LSTMs and GRUs have been shown to outperform feedforward DNNs \cite{sak2014sequence}. Later, a cascade of convolutional, LSTM and feedforward layers, i.e. the convolutional, long short-term memory deep neural network (CLDNN) model, 
was further shown to outperform LSTM-only models \cite{sainath2015convolutional}. In CLDNNs, a window of input frames is first processed by two convolutional layers with max-pooling layers to reduce the frequency variance in the signal, then projected down to a lower-dimensional feature space for the following LSTM layers to model the temporal correlations, and finally passed through a few feedforward layers and an output softmax layer.

With the adoption of RNNs for speech modeling, the conditional independence assumption of the output targets incurred by the traditional HMM-based phone state modeling is no longer necessary and the research field shifted towards full sequence-to-sequence models. There has been large interest in learning a purely neural sequence-to-sequence model, such as CTC and LAS. In~\cite{SoltauLiaoSak16}, Soltau et al.\ trained a CTC-based model with word output targets, which was shown to outperform a state-of-the-art CD-phoneme baseline on a YouTube video captioning task. The listen, attend and spell (LAS) model is a single neural network that includes an \emph{encoder} which is analogous to a conventional acoustic model, an \emph{attention module} that acts as an alignment model, and a \emph{decoder} that is analogous to the language model in a conventional system. 
Despite the architectural simplicity and empirical performance of such sequence-to-sequence models, further improvements in both model structure and optimization process have been proposed to outperform conventional models \cite{seq2seq_system2018}.

With dramatic improvements in speech recognition performance, it is robust enough for real world applications. Virtual assistants, such as Google Home, Amazon Alexa and Microsoft Cortana, all adopt voice as the main interaction modality. Speech transcriptions also find their way to various applications for retrieving information from multimedia, such as YouTube speech captioning. With increasing adoption of speech based applications, extending speech support for more speakers and languages has become more important.
Transfer learning has been used to boost the performance of ASR systems on low resource languages with data from rich resource languages \cite{li2018bytes}. With the success of deep learning models in ASR, other speech related tasks also embraces deep learning techniques, such as voice activity detection \cite{shuoyiinendpoint2017}, speaker recognition \cite{ravanelli2018speaker}, language recognition \cite{lopez2014automatic} and speech translation \cite{bansal2017towards}.

\subsubsection{Music}
\label{subsec:analysis:music}
Compared to speech, music recordings typically contain a wider variety of sound sources of interest.
In many kinds of music, their occurrence follows common constraints in terms of time and frequency, creating complex dependencies within and between sources.
This opens up a wide set of possibilities for automatic description of music recordings.

Tasks encompass
low-level analysis (onset and offset detection, fundamental frequency estimation),
rhythm analysis (beat tracking, meter identification, downbeat tracking, tempo estimation),
harmonic analysis (key detection, melody extraction, chord estimation),
high-level analysis (instrument detection, instrument separation, transcription, structural segmentation, artist recognition, genre classification, mood classification)
and high-level comparison (discovery of repeated themes, cover song identification, music similarity estimation, score alignment). 
Each of these has originally been approached with hand-designed algorithms or features combined with shallow classifiers, but is now tackled with deep learning.
Here a few chosen examples are highlighted, covering various tasks and methods.
Please refer to \cite{Bayle-github-2018} for a more extensive list.

Several tasks can be framed as binary event detection problems.
The most low-level one is onset detection, predicting which positions in a recording are starting points of musically relevant events such as notes, without further categorization.
It saw the first application of neural networks to music audio:
In 2006, Lacoste and Eck \cite{Lacoste2006_mlp} trained
a small MLP
on 200\,ms-excerpts of a constant-Q log-magnitude spectrogram to predict whether there is an onset in or near the center.
They obtained better results than existing hand-designed methods, and better than using an STFT, and observed no improvement from including phases.
Eyben et~al.\ \cite{Eyben2010_rnn} improved over this method,
 applying a bidirectional LSTM to spectrograms processed with a time difference filter, albeit using a larger dataset for training. 
Schlüter et~al.\ \cite{Schlueter2014_icassp} further improved results with a CNN processing 15-frame log-mel excerpts of the same dataset.
Onset detection used to form the basis for beat and downbeat tracking \cite{McFee2014_onsetbeats}, but recent systems tackle the latter more directly.
Durand et~al.\ \cite{Durand2016_downbeatcnn} apply CNNs and  Böck et~al.\ \cite{Boeck2016_beatdownbeatrnn}
train an RNN on spectrograms to directly track beats and downbeats.
Both studies rely on additional post-processing with a temporal model ensuring longer-term coherence than captured by the networks, either in the form of an HMM \cite{Durand2016_downbeatcnn} or Dynamic Bayesian Network (DBN) 
\cite{Boeck2016_beatdownbeatrnn}.
Fuentes et~al.\ \cite{Fuentes2018_downbeatcrnn} propose a CRNN that does not require post-processing, but also relies on a beat tracker.
A higher-level event detection task is to predict boundaries between musical segments. Ullrich et~al.\ \cite{ullrich_boundary_2014} solved it with a CNN, using a receptive field of up to 60\,s on strongly downsampled spectrograms.
Comparing approaches, both CNNs with fixed-size temporal context and RNNs with potentially unlimited context are used successfully for event detection. Interestingly, for the former, it seems critical to blur training targets in time \cite{Lacoste2006_mlp,Schlueter2014_icassp,ullrich_boundary_2014}.

An example for a multi-class sequence labelling problem is chord recognition, the task of assigning each time step in a (Western) music recording a root note and chord class.
Typical hand-designed methods rely on folding multiple octaves of a spectral representation into a 12-semitone \emph{chromagram} \cite{purwins_new_2000}, 
smoothing in time, and matching against predefined chord templates.
Humphrey and Bello \cite{Humphrey2012_cnn} note the resemblance to the operations of a CNN, and demonstrate good performance with a CNN trained on constant-Q, linear-magnitude spectrograms preprocessed with contrast normalization and augmented with pitch shifting.
Modern systems integrate temporal modelling, and extend the set of distinguishable chords.
As a recent example, McFee and Bello \cite{McFee2017_chords} apply a CRNN (a 2D convolution learning spectrotemporal features, followed by a 1D convolution integrating information across frequencies, followed by a bidirectional GRU) 
and use side targets to incorporate relationships between a detailed set of 170 chord classes.
Taking a different route, Korzeniowski et~al.\ \cite{Korzeniowski2016_deepchroma} train CNNs on log-frequency spectrograms to not only predict chords, but derive an improved chromagram representation useful for tasks beyond chord estimation.

Regarding sequence classification, one of the lowest-level tasks is to estimate the global tempo of a piece.
A natural solution is to base it on beat and downbeat tracking: downbeat tracking may integrate tempo estimation to constrain downbeat positions \cite{Boeck2016_beatdownbeatrnn,Durand2016_downbeatcnn}. 
However, just as beat tracking can be done without onset detection, Schreiber and Müller \cite{Schreiber2018_tempocnn} showed that CNNs can be trained to directly estimate the tempo from 12-second spectrogram excerpts, achieving better results and allowing to cope with tempo changes or drift within a recording.
As a broader sequence classification task encompassing many others, tag prediction aims to predict which labels from a restricted vocabulary users would attach to a given music piece.
Tags can refer to the instrumentation, tempo, genre, and others, but always apply to a full recording, without timing information.
Bridging the gap from an input sequence to global labels has been approached in different ways, which are instructive to compare.
Dieleman et~al.\ \cite{dieleman_end--end_2014} train a CNN with short 1D convolutions (i.e., convolving over time only) on 3-second log-mel spectrograms, and averaged predictions over consecutive excerpts to obtain a global label.
For comparison, they train a CNN on raw samples, with the first-layer filter size chosen to match typical spectrogram frames, but achieve worse results.
Choi et~al.\ \cite{Choi2016_tagging} use a FCN of $3\!\times\!3$ convolutions interleaved with max-pooling such that a 29-second log-mel spectrogram is reduced to a $1\!\times\!1$ feature map and classified.
Compared to FCNs in computer vision which employ average pooling in later layers of the network, max-pooling was chosen to ensure that local detections of vocals are elevated to global predictions.
Lee et~al.\ \cite{Lee2017_samplecnn} train a CNN on raw samples, using only short filters (size 2 to 4) interleaved with max-pooling, matching the performance of log-mel spectrograms.
Like Dieleman et~al., they train on 3-second excerpts and average predictions at test time.

To summarize, deep learning has been applied successfully to numerous music processing tasks, and drives industrial applications with automatic descriptions for browsing large catalogues, with content-based music recommendations in the absence of usage data, and also profanely with automatically derived chords for a song to play along with.
However, on the research side, neither within nor across tasks is there a consensus on what input representation to use (log-mel spectrogram, constant-Q, raw audio) and what architecture to employ (CNNs or RNNs or both, 2D or 1D convolutions, small square or large rectangular filters), leaving numerous open questions for further research.

\subsubsection{Environmental Sounds} 
\label{subsec:analysis:environmental}

In addition to speech and music signals, other sounds also carry a wealth of relevant information about our environments. Computational analysis of environmental sounds has several applications, for example in context-aware devices, acoustic surveillance, or multimedia indexing and retrieval.
It is typically done with three basic approaches: a) acoustic scene classification, b) acoustic event detection, and c) tagging.

Acoustic scene classification aims to label a whole audio recording with a single scene label. Possible scene labels include for example "home", "street", "in car", "restaurant", etc. The set of scene labels is defined in advance, rendering this a multinomial classification problem. Training material should be available from each of the scene classes. 

Acoustic event detection aims to estimate the start and end times of individual sound events such as footsteps, traffic light acoustic signalling, dogs barking, and assign them an event label. The set of possible event classes should be defined in advance. A simple and efficient way to apply supervised machine learning to do detection is to predict the activity of each event class in short time segments using a supervised classifier.
Usually, the supervised classifier used to do detection will use contextual information, i.e., acoustic features computed from the signal outside the segment to be classified.
A simple way to do so is to concatenate acoustic features from multiple context frames around the target frame, as done in the baseline method for the public DCASE (Detection and Classification of Acoustic Events and Scenes) evaluation campaign in 2016 \cite{mesaros-TASLP-2018}.
Alternatively, classifier architectures which model temporal information may be used: for example, recurrent neural networks may be applied to map a sequence of frame-wise acoustic features to a sequence of binary vectors representing event class activities \cite{PARASCANDOLO-ICASSP-2016}. Similarly to other supervised learning tasks, convolutional neural networks can be highly effective, but in order to be able to output an event activity vector at a sufficiently high temporal resolution, the degree of max pooling or stride over time should not be too large -- if a large receptive field is desired, dilated convolution and dilated pooling can be used instead \cite{Sercu2016_dilated}.

Tagging aims to predict the activity of multiple (possibly simultaneous) sound classes, without temporal information. In both tagging and event detection, multiple event classes can be targeted  that can be active simultaneously. In the context of event detection, this is called \emph{polyphonic} event detection. In this approach, the activity of each class can be represented by a binary vector where each entry corresponds to each event class, ones represent active classes, and zeros inactive classes. If overlapping classes are permitted, the problem is a \emph{multilabel} classification problem, where more than one entry in the binary vector can have value one.
 
It has been found out that using a multilabel classifier to jointly predict the activity of multiple classes at once produces better results, instead of using a single-class classifiers for each class separately. This might be for example due to the multiclass classifier being able to model the interaction of simultaneously active classes.

Since the analysis  of environmental sounds is a less established research field in comparison to speech and music, the size and diversity of available datasets for developing systems is more limited in comparison to speech and music datasets. Most of the open data has been published in the context of annual DCASE challenges. Because of the limited size of annotated environmental datasets, \emph{data augmentation} is a commonly used technique in the field, and it has been found highly effective.

\subsubsection{Localization and Tracking}  
\label{subsubsec:analysis:tracking}

Multichannel audio allows for the localization and tracking of sound sources, i.e. determining their spatial locations, and tracking them over time and can, for example, be used as a part of a source separation or speech enhancement system to separate a source from the estimated source direction, or in a diarization system to estimate the activity of multiple speakers.

A single microphone array consisting of multiple microphones can be used to infer the direction of a sound source, either in the azimuth, or in both azimuth and elevation. By combining information from multiple microphone arrays, directions can be merged to obtain source locations. Given a microphone array signal from multiple microphones, direction estimation can be formulated in two ways: 1) by forming a fixed grid of possible directions, and by using multilabel classification to predict if there is an active source in a specific direction \cite{Chakrabarty-NIPS-2017}, or 2) by using regression to predict the directions \cite{ferguson-ICASSP-2018} or spatial coordinates \cite{Vesperini-MLSP-2016} of target sources. In addition to this categorization, differences in various deep learning methods for localization lie in the input features used, the network topology, and whether one or more sources are localized. 

Commonly used input features that have been used for deep learning based localization include phase spectrum \cite{Chakrabarty-NIPS-2017}, magnitude spectrum \cite{ADAVANNE-EUSIPCO-2018}, and generalized cross-correlation between channels \cite{Vesperini-MLSP-2016}. In general, source localization requires the use of interchannel information, which can also be \emph{learned} by a deep neural network with a suitable topology from within-channel features, for example by convolutional layers \cite{ADAVANNE-EUSIPCO-2018} where the kernels span multiple channels.

\subsection{Synthesis and Transformation}
\label{subsec:synthesis}
\subsubsection{Source Separation} \label{subsec:synthesis:separation}

Source separation is the process of extracting the signal corresponding to individual sources from a mixture of multiple sources; this is important in audio signal processing, since in realistic environments, often multiple sources are present which sum to a mixture signal,  negatively affecting downstream signal processing tasks.
Example application areas related to source separation include music editing and remixing, preprocessing for robust classification of speech and other sounds, or preprocessing to improve speech intelligibility.

Source separation can be formulated as the process of extracting source signals $s_{m,i}(n)$ from the acoustic mixture
\begin{equation}
x_m(n) = \sum_{i=1}^I s_{m,i}(n),
\end{equation} where $i$ is the source index, $I$ is the number of sources, and $n$ is the sample index. In general, multiple microphones may be used to capture the audio, in which case $m$ is the microphone index and $s_{m,i}(n)$ is the spatial image of $i$th source in microphone $m$.

State-of-the-art source separation methods typically take the route of estimating \emph{masking} operations in the time-frequency domain (even though there are approaches that operate directly on time-domain signals and use a DNN to learn a suitable representation from it, see e.g. \cite{Pandey-INTERSPEECH-2018}). The reason for time-frequency processing stems mainly from three factors: 1) the structure of natural sound sources is more prominent in the time-frequency domain, which allows modeling them  more easily than time-domain signals, 2) convolutional mixing which involves an acoustic transfer function from a source to a microphone which can be approximated as instantaneous mixing in the frequency domain, simplifying the processing, and 3) natural sound sources are \emph{sparse} in the time-frequency domain which facilitates their separation in that domain.

Masking in the time-frequency domain may be formulated as a  multiplication of the mixture signal spectrum $X_m(f,t)$ at time $t$ and frequency $f$ by a separation mask $M_{m,i}(f,t)$ to obtain an estimate of the separated source signal spectrum of the $i$th source in the $m$th microphone channel as 
\begin{equation}
\hat{S}_{m,i}(f,t) = M_{m,i}(f,t) X_m(f,t).
\end{equation}

The spectrum $X_m(f,t)$ is typically calculated using the short-time-Fourier transform (STFT) because it can be implemented efficiently using the fast Fourier transform algorithm, and also because the STFT can be easily inverted.  The use of other time-frequency representations is also possible, such as constant-Q or mel spectrograms. The use of these has however become less common since they reduce output quality, and deep learning does not require a compact input representation that they would provide in comparison to the STFT.

Deep learning approaches operating on only one microphone rely on modeling the spectral structure of sources. They can be roughly divided in two categories: 1) methods that aim to predict the separation mask $M_{i}(f,t)$ based on the mixture input $X(f,t)$ (here the microphone index is omitted, since only one microphone is assumed), and 2) methods that aim to predict the source signal spectrum ${S}_{i}(f,t)$ from the mixture input. Deep learning in these cases is based on supervised learning based on the relation between  the input mixture spectrum $X(f,t)$ and the target output as either the \emph{oracle mask} or the clean signal spectrum \cite{WANG-TASLP-2014}. The oracle mask takes either binary values, or continuous values between 0 and 1. Various deep neural network architectures are applicable in the above settings, including the use of standard methods such as convolutional \cite{FU_2018} 
and recurrent \cite{HUANG-TASLP-2015} layers. The conventional mean-square error loss is not optimal for subjective separation quality, and therefore custom loss functions have been developed to improve intelligibility \cite{KOLBAEK-ICASSP-2018}.

A recent approach based on deep clustering \cite{ISIK-INTERSPEECH-2016} uses supervised deep learning to estimate embedding vectors for each time-frequency point, which are then clustered in an unsupervised manner. This approach allows separation of sources that were not present in the training set.
This approach can be further extended to a deep attractor network, which is based on estimating a single attractor vector for each source, and has been used to obtain  state-of-the-art results in single-channel source separation \cite{CHEN-ICASSP-2017}.

When multiple audio channels are available, e.g.\ captured by multiple microphones, the separation can be improved by taking into account the spatial locations of sources or the mixing process. In the multi-channel setting, a few different approaches exist that use deep learning. The most common approach is to use deep learning applied in a similar manner to single-channel methods, i.e.\ to model the single-channel spectrum or the separation mask of a target source \cite{Nugraha-TASLP-2016}; in this case the main role of deep learning is to model the spectral characteristics of the target. However, in the case of multichannel audio, the input features to a deep neural network can include spatial features in addition to spectral features (e.g. \cite{LIU-ICASSP-2018}). Furthermore, DNNs can be used to estimate the weights of a multi-channel mask (i.e., a beamformer)\cite{XIAO-ICASSP-2016}.  

Regarding the different audio domains, in speech it is assumed that the signal is sparse 
and that different sources are independent from each other. In environmental sounds, independence can usually be assumed. In music there is a high dependence between simultaneous sources as well as there are specific temporal dependencies across time, in the waveform as well as regarding long-term structural repetitions.

\subsubsection{Audio Enhancement}
\label{subsec:synthesis:enhancement}  
\label{sec:audio_enhancement}

Speech enhancement techniques aim to improve the quality of speech by reducing noise.
They are crucial components, either explicitly~\cite{chen2008fundamentals} or implicitly~\cite{xu2014experimental, li2016neural}, in ASR systems for noise robustness.
Besides conventional enhancement techniques~\cite{chen2008fundamentals}, deep neural networks have been widely adopted to either directly reconstruct clean speech~\cite{feng2014speech, xu2015regression} or estimate masks~\cite{wang2017supervised, li2014spectral, narayanan2013ideal} from the noisy signals.
 Conventional denoising approaches, such as Wiener methods, usually assume stationary noise, whereas deep learning approaches can model time-varying noise.
Different types of networks have been investigated in the literature for enhancement, such as denoising autoencoders~\cite{lu2013speech}, convolutional networks~\cite{FU_2018} 
and recurrent networks~\cite{weninger2015speech}.

Recently, GANs have been shown to perform well in speech enhancement in the presence of additive noise~\cite{pascual2017segan}, when enhancement is posed as a translation task from noisy signals to clean ones. The proposed speech enhancement GAN (SEGAN) yields improvements in perceptual speech quality metrics over the noisy data and a traditional enhancement baseline. In \cite{donahue2018exploring}, GANs are used to enhance speech represented as log-mel spectra. When GAN-enhanced speech is used for ASR, no improvement is found compared to enhancement using a simpler regression approach.

\subsubsection{Generative Models} 
\label{subsec:synthesis:generative}

Generative sound models synthesize sounds according to characteristics learned from a sound data-base, yielding realistic sound samples.  The generated sound should be  similar to  sounds from which the model is trained, in terms of typical acoustic features (timbre, pitch content, rhythm). A basic requirement is that the sound  should be recognizible as stemming from a particular object/process or intelligible, in the case of speech generation.  At the same time,  the generated sound should be original, i.e.~it should be significantly different from sounds in the training set, instead of simply copying training set sounds. A further requirement is that the generated sounds should show diversity. 
It is desirable to condition the sound synthesis, e.g.  in speech synthesis on a speaker, a prosodic trajectory, a harmonic schema in music, or physical parameters in the generation of environmental sounds.  In addition, training and generation time should be small; ideally generation should be possible in real-time. Sound synthesis may be performed based on a spectral representation (e.g. log-mel spectrograms) or from raw audio. The former representation lacks the phase information that needs to be reconstructed in the synthesis,
 e.g.\ via the Griffin-Lim algorithm \cite{daniel_w._griffin_signal_1984} in combination with the inverse Fourier transform \cite{wang_tacotron2017} 
which does not reach high synthesis quality. End-to-end synthesis may be performed block-wise or with an autoregressive model, where sound is generated sample-by-sample, each new sample conditioned on previous samples. In the blockwise approach, in the case of variational autoencoder (VAE)  or GANs  \cite{donahue_synthesizing_2018}, the sound is often synthesised from a low-dimensional latent representation, from which it needs to by upsampled (e.g. through nearest neighbor or linear interpolation) to the high resolution sound. Artifacts, induced by the different layer resolutions, can be ameliorated 
through random phase perturbation in different layers \cite{donahue_synthesizing_2018}. 
 In the autoregressive approach,  the new samples are  synthesised iteratively, based on an infinitely long context of  previous samples, when using RNNs (such as LSTM or GRU), at the cost of expensive computation when training. However, layers of RNNs may be stacked to process the sound on different temporal resolutions, where the activations of one layer depend on the activations of the next layer with coarser resolution \cite{mehri_samplernn2016}. An efficient audio generation model \cite{kalchbrenner_efficient_2018} based on sparse RNNs folds long sequences into a batch of shorter ones.  Stacking dilated convolutions in the WaveNet \cite{oord2016wavenet}  can lead to context windows of reasonable size. Using WaveNet \cite{oord2016wavenet}, the autoregressive sample prediction is cast as a classification problem, the amplitude of the predicted sample being quantized logarithmically into distinct classes, each corresponding to an interval of amplitudes. 
 Containing the  samples, the input can be extended with context information \cite{oord2016wavenet}.
   This context may be global (such as a speaker identity) or  changing during time (such as f0 or mel spectra)\cite{oord2016wavenet}. In \cite{shen2017natural}, a text-to-speech system is introduced which consists of two modules: (1) a neural network is trained from textual input to predict a sequence of mel spectra, used as contextual input to (2) a WaveNet yielding synthesised speech. 
    WaveNet-based models 
    for speech synthesis outperform state-of-the-art systems by a large margin, but their training is computationally expensive. The development of parallel WaveNet \cite{oord2017parallel} provides a solution to the slow training problem and hence speeds up the adoption of WaveNet models in other applications 
    \cite{shen2017natural,chen2018high,chang2018temporal}.
      In \cite{engel_neural2017}, synthesis is controlled through parameters in the latent space of an autoencoder, applied e.g. to morph between different instrument timbres.
      Briot et al. \cite{briot_deep_2017} provide a more in-depth treatment of music generation with deep learning.

Generative models can be evaluated both objectively or subjectively: Recognizability of generated sounds can be tested objectively through a classifier (e.g. inception score 
in \cite{donahue_synthesizing_2018}) or subjectively in a forced choice test with humans. Diversity can be objectively assessed. Sounds being represented as normalized log-mel spectra, diversity can be measured as  the average Euclidean distance between
the sounds and their nearest neighbors. Originality can be measured as the  average Euclidean distance between a generated samples to their nearest neighbor in the real training set  \cite{donahue_synthesizing_2018}. 
 A Turing test, asking a human to distinguish between real and synthesized audio examples, is a hard test for a model, since passing the Turing test requires that there is no perceivable difference between an example being  real or being  synthesized. The WaveNet, for example, yields a higher MOS than concatenative or parametric methods, which represented the previous state of the art \cite{oord2016wavenet}. 

\section{Discussion and Conclusion} 
\label{sec:discussion}
\vspace{-0.02in}

In this section, we look at deep learning across the different audio domains, regarding the following aspects: features (Sec.~\ref{subsec:discussion:features}),
models (Sec.~\ref{subsec:discussion:models}), data requirements (Sec.~\ref{subsec:discussion:data}),  computational complexity (Sec.~\ref{subsec:discussion:complexity}), interpretability and adaptability (Sec.~\ref{subsec:discussion:interpretability}).
For each aspect, we highlight differences and similarities between the domains, and note common challenges worthwhile to work on.

\subsection{Features}
\label{subsec:discussion:features}
Whereas MFCCs are the most common representation in traditional audio signal processing, log-mel spectrograms are the dominant feature in deep learning, followed by raw waveforms or complex spectrograms. Raw waveforms avoid hand-designed features, which should allow to better exploit the improved modeling capability of deep learning models, learning representations optimized for a task. However, this incurs higher computational costs and data requirements, and benefits may be hard to realize in practice.   
For analysis tasks, such as ASR, MIR, or environmental sound recognition, log-mel spectrograms provide a more compact representation, and methods using these features usually need less data and training to achieve results that are, at the current state of the art, comparable in classification performance to a setup where raw audio is used.
In a task where the aim is to  synthesize a sound of high audio quality, such as in source separation, audio enhancement, TTS, or sound morphing, using (log-mel) magnitude spectrograms poses the challenge to reconstruct the phase. In that case, raw waveforms or complex spectrograms are generally preferred as the input representation.

However, some works report improvements using raw waveforms for analysis tasks \cite{Ghahremani2016_rawaudio,Sailor2016_rawaudio,oord2016wavenet}, and some attempt to find a way in between by designing and/or initializing the first layers of a deep learning system to mimic engineered representations \cite{sainath_learning_2013,Cakir2016_stftfilterbank,Yedid15,Sainath15b}.
So there are still several open research questions:
Are mel spectrograms indeed the best representation for audio analysis?
Under what circumstances is it better to use the raw waveform?
Can we do better by exploring the middle ground, a spectrogram with learnable hyperparameters?
If we learn a representation from the raw waveform, does it still generalize between tasks or domains?

\subsection{Models}
\label{subsec:discussion:models}

 On a historical note, in ASR, MIR, and environmental sound analysis, deep models have replaced support vector machines for sequence classification, and   GMM-HMMs for  sequence transduction.  
In audio enhancement / denoising and source separation, deep learning has solved tasks previously addressed by non-negative matrix factorization and Wiener methods, respectively.
In audio synthesis, concatenative synthesis has been replaced e.g. by Wavenet, SampleRNN, WaveRNN.

Across the domains, CNNs, RNNs and CRNNs are employed successfully, with no clear preference.
All three can model temporal sequences, and solve sequence classification, sequence labelling and sequence transduction tasks.
CNNs have a fixed receptive field, which limits the temporal context taken into account for a prediction, but at the same time makes it very easy to widen or narrow the context used.
RNNs can, theoretically, base their predictions on an unlimited temporal context, but first need to learn to do so, which may require adaptations to the model (such as LSTM) and prevents direct control over the context size.
Furthermore, they require processing the input sequentially, making them slower to train and evaluate on modern hardware than CNNs.
CRNNs offer a compromise in between, inheriting both CNNs and RNNs advantages and disadvantages.

Thus, it is an open research question which model is superior in which setting.
From existing literature, this is very hard to answer, since different research groups yield state-of-the-art results with different models.
This may be due to each research group's specialized informal knowledge about how to effectively design and tune a particular architecture type.

\subsection{Data Requirements}
\label{subsec:discussion:data}

With the possible exception of speech recognition, in industry, for the most widespread languages, all tasks in all audio domains face relatively small datasets, posing a limit on the size and complexity of deep learning models trained on them.

In computer vision, a shortage of labeled data for a particular task is offset by the widespread availability of models trained on the ImageNet dataset \cite{imagenet}:
To distinguish a thousand object categories, these models learned transformations of the raw input images that form a good starting point for many other vision tasks.
Similarly, in neural language processing, word prediction models trained on large text corpora have shown to yield good model initializations for other language processing tasks \cite{elmo,bert}.
However, no comparable task and dataset -- and models pretrained on it -- exists for the audio domain.

This leaves several research questions.
What would be an equivalent task for the audio domain?
Can there be an audio dataset covering speech, music, and environmental sounds, used for transfer learning, solving a great range of audio classification problems?
How may pre-trained audio recognition models be flexibly adapted to new tasks using a minimal amount of data, i.e. to out-of-vocabulary words, new languages, new musical styles and new acoustic environments?
It is well possible that this has to be answered separately for each domain, rather than across audio domains.
Even just within the music domain, while transfer learning might work for global labels like artists and genres, individual tasks like harmony detection or downbeat detection might be too different to transfer from one to another.

If transfer learning turns out to be the wrong direction for audio, research needs to explore other paradigms for learning more complex models from scarce labeled data,   such as semi-supervised learning, active learning, or few-shot learning.

\subsection{Computational Complexity}
\label{subsec:discussion:complexity}

The success of deep neural networks leverages the advances of fast and large scale computations. Compared to conventional approaches, state-of-the-art deep neural networks usually require more computation power and more training data.
CPUs are not optimally suited for training and evaluating large deep models. Instead, processors optimized for matrix operations are commonly used, mostly general-purpose graphics processing units (GPGPUs) \cite{che2008performance} and application-specific integrated circuits such as the proprietary tensor processing units (TPUs) \cite{jouppi2017datacenter}.

Applications with strict limits on computational resources, such as mobile phones or hearing instruments, require smaller models.
While a lot of recent works tackle the simplification, compression or training of neural networks with minimal computational budgets, it may be worthwhile to explore options for the specific requirements of real-time audio signal processing.

\subsection{Interpretability and Adaptability}
\label{subsec:discussion:interpretability}  
 
In deep learning, researchers usually design a network structure using primitive layer blocks and a loss function for the target task. The parameters of the model are learned by gradient descent on the loss for pairs of inputs and targets or inputs only for unsupervised training. The connection between the layer parameters and the actual task is hard to interpret.  Researchers have been attempting to relate the activities of the network neurons to the target tasks (e.g., \cite{Schlueter2014_icassp,tan2015improving}), or investigate which parts of the input a prediction is based on (e.g., \cite{schlueter2016_ismir,Mishra2017_slime}). Further research into understanding how a network or a sub network behaves could help improving the model structure to address failure cases.

\ifCLASSOPTIONcaptionsoff
  \newpage
\fi




\bibliographystyle{IEEEtran}
\bibliography{DL4AudioReviewJSTSP}
\end{document}